\documentclass[amsmath, amssymb, aps, longbibliography, prl, superscriptaddress,
preprint,
nobibnotes, 
endfloat,
]{revtex4-2}

\usepackage[english]{babel}
\makeatletter\expandafter\let\csname l@en\endcsname\l@english\makeatother

\makeatletter
\def\NAT@spacechar{}%   \hspace{.5ex}
\makeatother

\usepackage{amssymb}
\usepackage{amsfonts}
\usepackage{amsmath}
\usepackage[separate-uncertainty=true]{siunitx}
\usepackage[colorlinks=true]{hyperref}% add hypertext capabilities
\AtBeginDocument{\definecolor{linkcolor}{rgb}{0,0.4470,0.7410}\hypersetup{allcolors=linkcolor}}
\usepackage{graphicx}% Include figure files

\let\oldfrac\frac% Store \frac
\renewcommand{\frac}[2]{%
  \mathchoice
    {\oldfrac{#1}{#2}}% display style
    {#1/#2}% text style
    {#1/#2}% script style
    {#1/#2}% script-script style
}

\let\oldautoref\autoref{}

\renewcommand{\autoref}[1]{%
    \begingroup%
    \def\figureautorefname{Figure}%
    \def\tableautorefname{Table}%
    \def\partautorefname{Part}%
    \def\appendixautorefname{Appendix}%
    \def\equationautorefname{Equation}%
    \def\itemautorefname{Item}%
    \def\chapterautorefname{Chapter}%
    \def\sectionautorefname{Section}%
    \def\subsectionautorefname{Subsection}%
    \def\subsubsectionautorefname{Subsubsection}%
    \def\paragraphautorefname{Paragraph}%
    \def\Hfootnoteautorefname{Footnote}%
    \def\AMSautorefname{Equation}%
    \def\theoremautorefname{Theorem}%
    \oldautoref{#1}%
    \endgroup%
}

\newlength{\plotwidth}

\newcommand{\gtwo}{$g^{(2)}$}

\begin{document}
%\preprint{APS/123-QED}

\title{Fluorescence intensity correlations enable 3D imaging without sample rotations}% Force line breaks with \\

\newcommand{\UHH}{\affiliation{Center for Free-Electron Laser science (CFEL), Universit\"at Hamburg, 22761 Hamburg, Germany}}
\newcommand{\CHYN}{\affiliation{Center for Hybrid Nanostructures (CHyN), Universit\"at Hamburg, 22761 Hamburg, Germany}}
\newcommand{\LCLS}{\affiliation{Linac Coherent Light Source, SLAC National Accelerator Laboratory, Menlo Park, CA 94025, USA}}
\newcommand{\SLAC}{\affiliation{SLAC National Accelerator Laboratory, Menlo Park, CA 94025, USA}}
\newcommand{\PULSE}{\affiliation{Stanford PULSE Institute, SLAC National Accelerator Laboratory, Menlo Park, CA 94025, USA}}
\newcommand{\Stanford}{\affiliation{Physics Department, Stanford University, Stanford, CA 94025, USA}}
\newcommand{\StanfordApplied}{\affiliation{Department of Applied Physics, Stanford University, Stanford, CA, USA}}
\newcommand{\Argonne}{\affiliation{Argonne National Laboratory, Argonne, IL 60439, USA}}
\newcommand{\PSI}{\affiliation{PSI Center for Photon Science, 5232 Villigen PSI, Switzerland}}
\newcommand{\TUD}{\affiliation{Technical University Darmstadt, Institute of Nuclear Physics, Schlossgartenstr. 9, 64289 Darmstadt, Germany}}
\newcommand{\SACLA}{\affiliation{RIKEN SPring-8 Center, 1-1-1 Kouto, Sayo, Hyogo 679-5148, Japan}}
\newcommand{\PTB}{\affiliation{Physikalisch Technische Bundesanstalt (PTB) Berlin and Braunschweig, Germany}}
\newcommand{\SZFKI}{\affiliation{Institute for Solid State Physics and Optics, Wigner Research Centre for Physics, Budapest, Hungary}}
\newcommand{\OSAKA}{\affiliation{Graduate School of Engineering, Osaka University, 2-1 Yamada-oka, Suita, Osaka 565-0871, Japan}}

\author{Robert~G.~Radloff}\thanks{These authors contributed equally to this work}\email[Corresponding author: ]{robert.radloff@cfel.de}\UHH{}
\author{Felix~F.~Zimmermann}\thanks{These authors contributed equally to this work}\UHH{}\PTB{}\SLAC{}
\author{Siqi~Li}\SLAC{}
\author{Stephan~Kuschel}\UHH{}\TUD{}
\author{Anatoli~Ulmer}\UHH{}
\author{Yanwen~Sun}\SLAC{}
\author{Takahiro~Sato}\SLAC{}
\author{Peihao~Sun}\SLAC{}
\author{Johann~Haber}\SLAC{}
\author{Diling~Zhu}\SLAC{}
\author{Miklós~Tegze}\SZFKI{}
\author{Gyula~Faigel}\SZFKI{}
\author{Matthew~R.~Ware}\SLAC{}
\author{Jordan~T.~O'Neal}\SLAC{}
%\author{Yoshiaki Kumagai}\SACLA{}
\author{Jumpei~Yamada}\OSAKA{}\SACLA{} 
\author{Taito~Osaka}\SACLA{}
\author{Robert~Zierold}\CHYN{}
\author{Carina~Hedrich}\CHYN{}
\author{Dimitrios Kazazis}\PSI{}
\author{Yasin~Ekinci}\PSI{}
\author{Makina~Yabashi}\SACLA{}
\author{Ichiro~Inoue}\SACLA{}
\author{Andrew~Aquila}\SLAC{}
\author{Meng~Liang}\SLAC{}
\author{Agostino~Marinelli}\SLAC{}
\author{Tais~Gorkhover}
\email[Corresponding author: ]{tais.gorkhover@cfel.de}\UHH{}

\date{\today}

%\input{chapters/abstract}
%\input{keywords}
%%%%%%%%%%%%%%%%%%%%%%%%%%%%%%%%%%%%%%%%%%%% Abstract %%%%%%%%%%%%%%%%%%%%%%%%%%%%%%%%%%%%%%%%%%%%%%%%%%%
\begin{abstract}
Lensless X-ray imaging provides element-specific nanoscale insights into thick samples beyond the reach of conventional light and electron microscopy. Coherent diffraction imaging (CDI) methods, such as ptychographic tomography, can recover three-dimensional (3D) nanoscale structures but require extensive sample rotation, adding complexity to experiments. X-ray elastic-scattering patterns from a single sample orientation are highly directional and provide limited 3D information about the structure. In contrast to X-ray elastic scattering, X-ray fluorescence is emitted mostly isotropically. However, first-order spatial coherence has traditionally limited nanoscale fluorescence imaging to single-crystalline samples. Here, we demonstrate that intensity correlations of X-ray fluorescence excited by ultrashort X-ray pulses contain 3D structural information of non-periodic, stationary objects. In our experiment, we illuminated a vanadium foil within a sub-200 nm X-ray laser beam focus. Without changing the sample orientation, we recorded 16 distinct specimen projections using detector regions covering different photon incidence angles relative to the X-ray free-electron laser (FEL) beam. The projections varied systematically as the fluorescing volume was translated along an astigmatism, confirming that FEL-induced fluorescence reflects real-space structural changes. Our results establish a new approach for lensless 3D imaging of non-periodic specimens using fluorescence intensity correlations, with broad implications for materials science, chemistry, and nanotechnology.
\end{abstract}

\maketitle

\section{Introduction}
X-ray imaging provides unique structural information of nanoscale features in matter beyond the capabilities of optical light, ions, and electrons. Hard X-rays can create precise three-dimensional electron density maps even inside {\textmu}m-thick and optically opaque specimens. Prominent applications include coherent diffraction imaging (CDI) schemes such as ptychographic tomography \cite{dierolf2010ptychographic, holler2014x, aidukas2024high} or Bragg CDI \cite{pfeifer2006three, ulvestad2015topological, vicente2021bragg}, where multiple projections from various specimen rotations can be assembled into full three-dimensional nanoscale models of samples ranging from computer chips to inner workings of batteries. CDI is based on coherent elastic X-ray scattering. The scattered X-rays interfere and produce distinct diffraction patterns along specific directions. Thus, a single orientation image provides a two-dimensional projection of the specimen and very limited three-dimensional structural insights \cite{Neutze2000potential, Seibert2011single, barke20153d}. 
In contrast, X-ray fluorescence emission is almost isotropic and thus removes the need to rotate the sample. Kossel line fitting \cite{gog1995kossel,bortel20243d} and X-ray fluorescence holography \cite{tegze1996x,Gog1996} are prominent examples of three-dimensional mapping through rotation of the detector. Until today, these principles have been applied exclusively to single-crystalline specimens due to the low degree of spatial coherence in X-ray fluorescence \cite{ma_structure_1994}. Recent studies have theoretically proposed that second-order coherence of X-ray fluorescence can be utilized to resolve structural maps of fluorescing nanoscale objects \cite{classen2017, ho2021fluorescence,lohse_incoherent_2021,shevchuk2021, trost2020photon,trost_speckle_2023}. To our knowledge, experimental studies have so far demonstrated only two-dimensional projections of non-periodic fluorescing volumes \cite{classen2017, nakamura_focus_2020, trost2023imaging}. Theoretical studies have predicted that intensity correlations of X-ray fluorescence, or incoherent diffraction imaging (IDI)\cite{classen2017}, can also be leveraged for 3D imaging capabilities of non-periodic specimens \cite{classen2017, ho2021fluorescence}.  If fluorescence is induced by ultrashort free-electron-laser (FEL) pulses, which are shorter than the natural lifetime of the fluorescent state \cite{Krause1979natural}, the X-ray fluorescence bursts are emitted within their coherence time. Such emissions can be shorter than temporal phase fluctuations and form a stationary interference pattern in the form of speckles. In this case, the nano-object is akin to a source of pseudothermal light, which is used in visible light experiments. 
In the present study, we demonstrate that intensity correlations of such fluorescence speckles carry three-dimensional tomographic information of a stationary and non-periodic sample volume. In our experiment, different tiles of our detectors resolved 16 distinct projections of a vanadium foil placed in a sub-200 nm FEL focus. The orientation of the foil relative to the FEL beam remained constant throughout our experiment. We have changed the shape of the fluorescing volume by shifting the foil along the laser beam direction, both inside and slightly outside the Rayleigh length. The changes in the structure of the fluorescing volume mirrored an astigmatism, which was not captured by the wavefront sensor during alignment. The observed projections on the individual tiles changed consistently with the expected geometry of an astigmatic Gaussian beam focus.
% 
%%%%%%%%%%%%%%%%%%%%  Figure 1: Principle  %%%%%%%%%%%%%%%%%%%%%%%%%%%%%%%%
\section{Tomographic character of X-ray fluorescence intensity correlations}
\begin{figure}[p]
    \centering
    \includegraphics[width=1\textwidth]{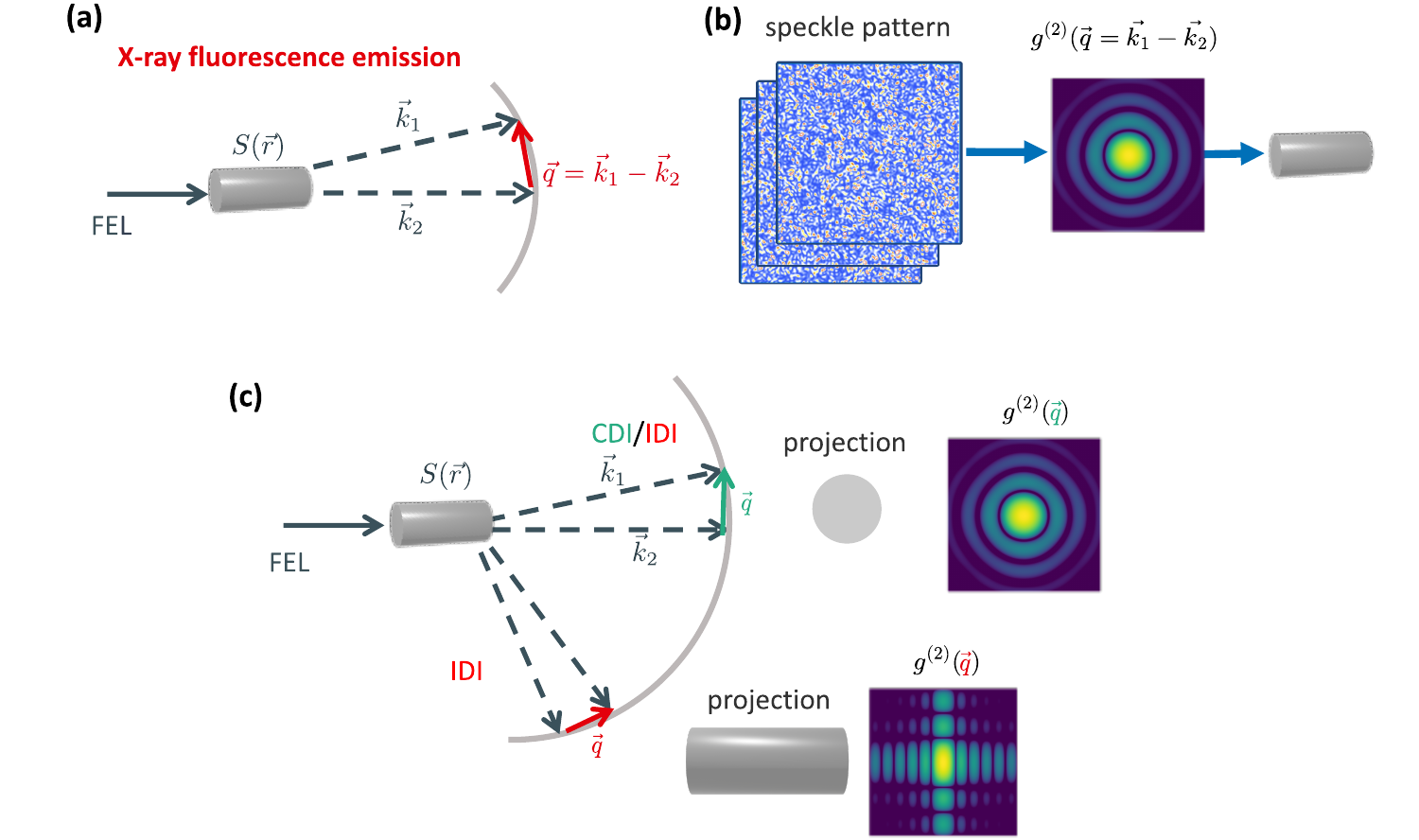} 
    \caption{
    Basic principle of incoherent diffractive imaging and its three-dimensional imaging capabilities.
    \textbf{a}~An object $S(\vec{r})$ is illuminated by the FEL pulse and subsequently emits X-ray fluorescence. The fluorescence photons are recorded on a detector (here indicated by the light-gray curve). The second-order intensity correlation is calculated between different pixels on the detector corresponding to different wavevectors $\vec{k}_1,\vec{k}_2$ of the detected photons.
    \textbf{b}~Due to the random phase relation between individual emitters, fluorescence produces speckle patterns. The structure factor of the emitting object $\left|\tilde{S}(\vec{q})\right|^2$ can be retrieved from these speckle patterns via the second-order intensity correlation $g^{(2)}(\vec{q})$. The retrieval of the structure factor $\left|\tilde{S}(\vec{q})\right|^2$ then allows the reconstruction via phase retrieval akin to the process in CDI.
    \textbf{c}~Opposed to CDI, IDI is not restricted by the scattering signal at high angles, as fluorescence is emitted isotropically. Hence, IDI can record several projections of the sample simultaneously, while CDI is limited to a single projection per sample orientation. 
    }
      \label{fig:principle}
\end{figure}

The tomographic character of IDI was proposed in the original paper by Classen et al. \cite{classen2017} and is sketched in \autoref{fig:principle}~(a). Here, a sample volume $S(\vec{r})$ at the real-space position $\vec{r}$ emits temporally coherent X-ray fluorescence. The fluorescence photons with wavevectors $\vec{k_1}$ and $\vec{k_2}$ of the wavelength $\lambda = \frac{2\pi}{|\vec{k}|}$ are registered simultaneously by a two-dimensional, far-field X-ray detector by different pixels. The fluorescence pulses will have a high degree of temporal coherence if the excitation X-ray pulse is short compared to the lifetime of the specific fluorescent state. For fluorescence in the hard X-ray regime, the coherence time is $\lesssim \qty{1}{\femto\second}$. Thus, newly available sub-fs to few fs Free-Electron-Laser (FEL) pulses \cite{yan2024, inoue2025} are of central importance for IDI studies. The high temporal coherence of the fluorescence leads to the formation of speckle patterns (presented in \autoref{fig:principle}~(b), left side) which contain information about the structure of the sample \cite{hanbury_brown_correlation_1956, hanbury_brown_test_1956, glauber1963a}. Intensity correlations averaged over multiple exposures provide the second-order correlation function $$g^{(2)}(\vec{k}_1,\vec{k}_2) = \frac{
    \left<I(\vec{k}_1) I(\vec{k}_2)\right>
}{
    \left<I(\vec{k}_1)\right>\left<I(\vec{k}_2)\right>
}$$
The Siegert relation
$
g^{(2)}\left(\vec{k}_{1}, \vec{k}_{2}\right)-1=\left|g^{(1)}\left(\vec{k}_{1} - \vec{k}_{2}\right)\right|^{2}
$
connects \gtwo{} with the normalized first-order correlation function $$g^{(1)}(\vec{k}_1,\vec{k}_2) = \frac{
    \left<E^*(\vec{k}_1) E(\vec{k}_2)\right>
}{
    \left[\left<\left|E(\vec{k}_1)\right|^2\right>\left<\left|E(\vec{k}_2)\right|^2\right>\right]^{\frac{1}{2}}
}$$ $E(\vec{k})$ and $E^*(\vec{k})$ are the amplitude of the electric field at a given $\vec{k}$ and its complex conjugate, respectively. Homogeneity of the fluorescence makes \gtwo{} shift invariant and only dependent on $\vec{q}=\vec{k_1}-\vec{k_2}$. 
For extended samples that emit more than two photons onto the detector per exposure, correlation signals from all pixel pairs corresponding to the same $\vec{q}$ can be averaged.
The volume of the accessible $\vec{q}$ space in IDI is distinct from that in CDI. In IDI, the highest accessible $\vec{q}$ spans the full detector, whereas in CDI it spans only half the detector at maximum. The $\vec{q}$ in IDI are not limited to the Ewald sphere as $\vec{k_1}$ and $\vec{k_2}$ can be freely chosen on the detector (\autoref{fig:principle}~(c)).
According to the van~Cittert-Zernike theorem, $g^{(1)}$ is proportional to the Fourier transform $\tilde{S}(\vec{q})$ of the fluorescence source $S(\vec{r})$ \cite{carter1981}.
Therefore, by measuring $\left|g^{(1)}\right|^2$ via the Siegert relation, we directly access the sample's Fourier magnitudes,$\left|\tilde{S}(\vec{q})\right|^2$. The real-space structure of $S(\vec{r})$ can then be recovered using standard phase retrieval algorithms \cite{fienup1982, elser2003}.
\section{Simultaneous measurement of multiple sample projections on a single detector}
\begin{figure}[p]
    \centering
    \includegraphics[width=1\textwidth]{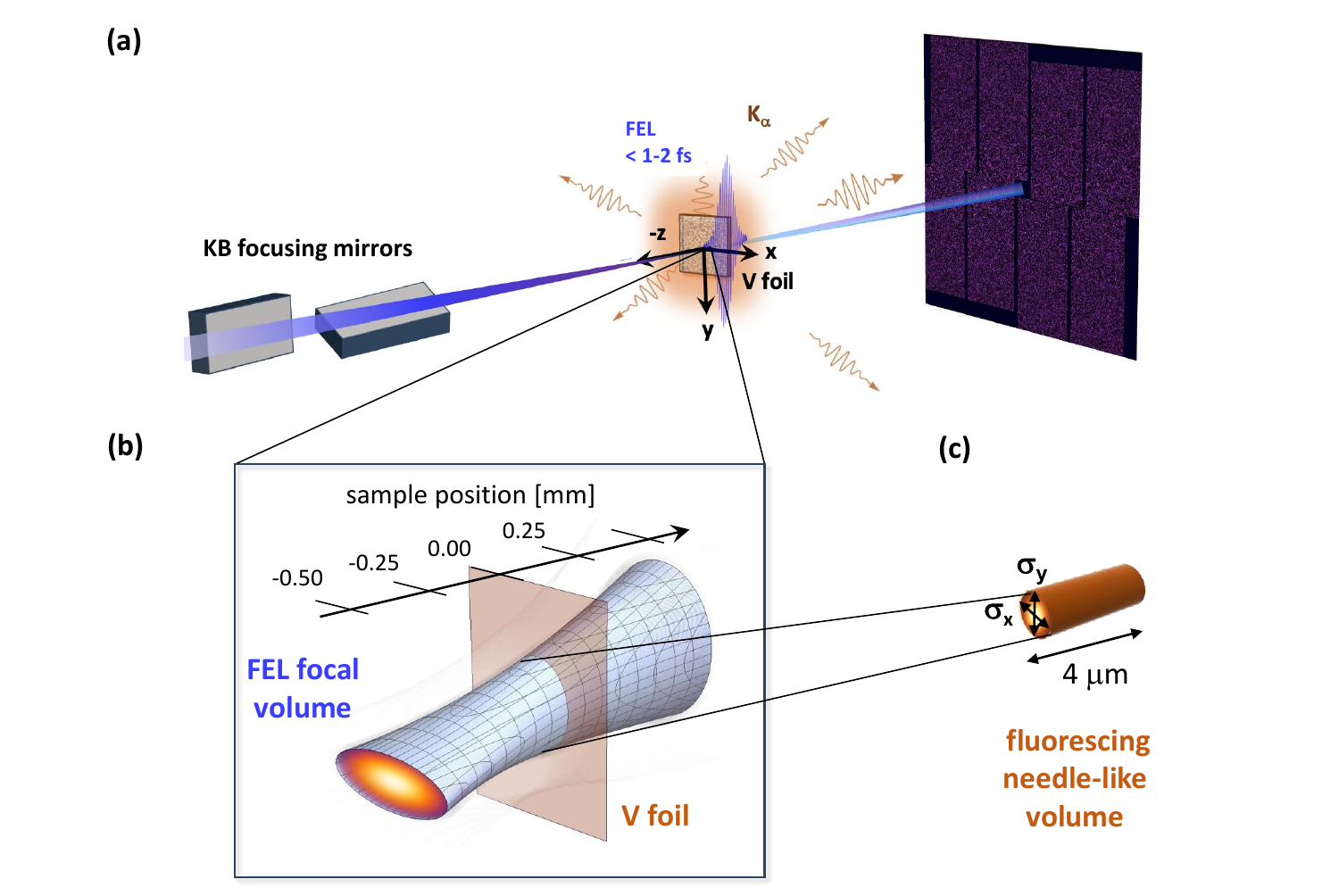} 
    \caption{
    Description of the experimental setup and visualization of the focus astigmatism.
    \textbf{a}~Schematic representation of the experimental setup at the CXI endstation at LCLS. The few-femtosecond X-ray pulses provided by LCLS were focused using CXI's KB mirror system to a sub-\qty{200}{\nano\meter} focus spot size. The {4{$\;$}{\textmu}m} thick vanadium foil was placed in the X-ray beam at varying distances from the focus spot. The emitted fluorescence was recorded on a Jungfrau~4M detector placed \qty{50}{\centi\meter} downstream from the sample.
    \textbf{b}~The focus spot has a different shape on the vanadium foil when the foil is shifted relative to the nominal focus position within an astigmatic focal volume.
    \textbf{c}~The fluorescing volume appears as a long needle from the perspective of the detector. Regions close to the center of the detector see a mostly two-dimensional Gaussian spot as the needle is viewed point-on. Regions closer to the edge of the detector see a more elongated volume because the needle is viewed from the side. 
    }
      \label{fig:experiment}
\end{figure}

\begin{figure}[p]
    \centering
    \includegraphics[width=1\textwidth]{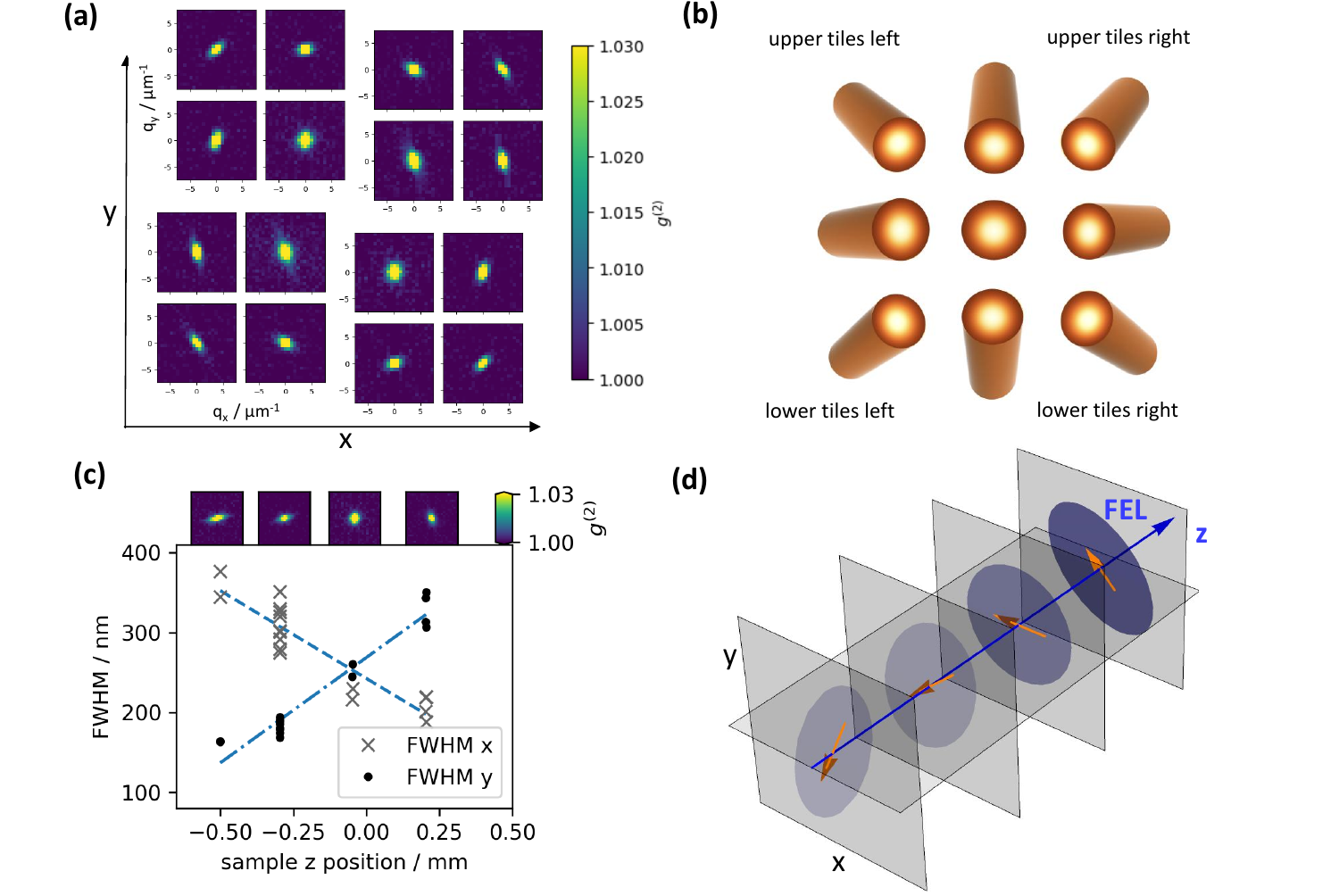} 
    \caption{
    Experimental results of the \gtwo{} measurement.
    \textbf{a}~Resulting \gtwo{} with the vanadium foil at the nominal focus position. Every Jungfrau~4M detector tile has a different perspective on the fluorescing volume.
    \textbf{b}~Schematic representation of the different perspectives onto the fluorescing needle as seen by the different detector tiles.
    \textbf{c}~Fitted focal spot size at different foil positions along the beam axis with a guide to the eye (blue lines). The insets at the top show the measured \gtwo{} at respective foil positions. The tile chosen for the fitting procedure was the upper-left center tile in \autoref{fig:results}~(a), which is the closest to the point-on perspective.
    \textbf{d} Simulated cross section of an astigmatic Gaussian beam at positions similar to those in the experiment. The orange arrows represent the tilt of the major axes of the ellipses.
    }
    \label{fig:results}
\end{figure}

\begin{figure}[p]
    \centering
    \includegraphics[width=1\textwidth]{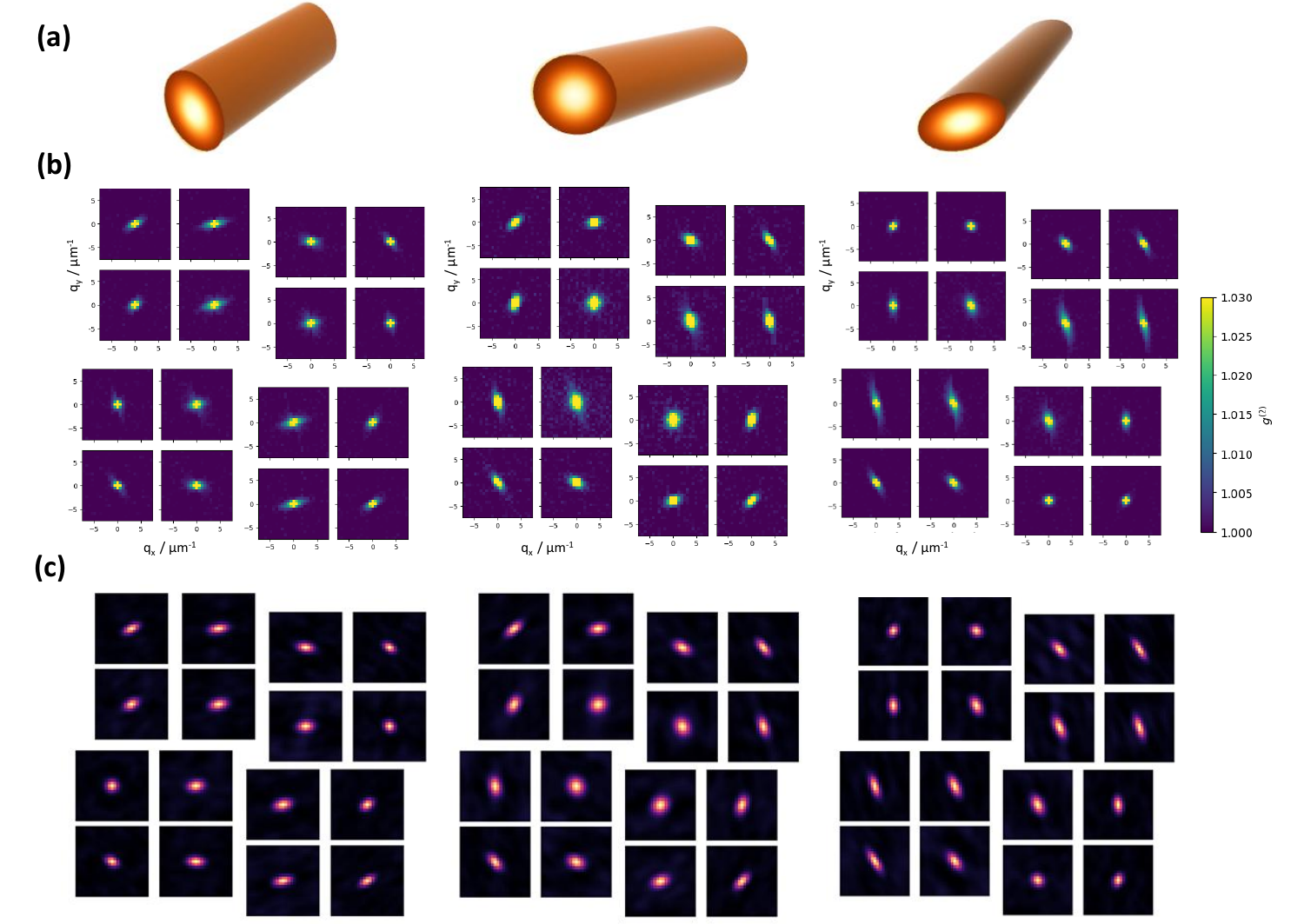} 
    \caption{
    Comparison between the measured and simulated \gtwo{} from the fluorescing foil volume at the nominal focus and \qty{250}{\micro\meter} upstream and downstream from the focus.
    \textbf{a}~A sketch of the fluorescing volume at the respective foil position shifted along the astigmatism of the FEL.
    \textbf{b}~Measured \gtwo{} upstream (left), downstream (right) of the nominal position and at the nominal focus (center).
    \textbf{c}~Simulated \gtwo{} at the same conditions as in the experiment displayed in (b). The simulation closely reproduces the experimental data, especially in the shape and defocusing effect along the astigmatism.
    }
      \label{fig:simulation}
\end{figure}

%%%%%%%%%  Figure 2: Experimental setup  %%%%%%%%%%%
The experiment inside the nanofocus chamber at the CXI endstation \cite{Liang2015coherent}  of the Linac Coherent Light Source (LCLS) \cite{Emma2010} is sketched in \autoref{fig:experiment}~(a). A pair of Kirkpatrick-Baez mirrors focused \qty{10.48}{\kilo\eV} X-ray pulses onto a \qty{4}{\micro\meter} thin vanadium foil (supplier Goodfellow, purity \qty{99.8}{\percent}). The resulting fluorescence was recorded by the Jungfrau~4M detector \cite{mozzanica2018jungfrau} placed \qty{50}{\centi\meter} further downstream. We minimized the pulse duration using non-linear compression to a few femtosecond durations with FEL pulse energies of around \qty{15}{\micro\joule} \cite{Huang2017generating}. A snake scan of the vanadium sample ensured that each FEL exposure illuminates a fresh spot on the foil. The optimal nanofocus spot at CXI is $< 200\;$nm full width half maximum (FWHM) in all directions. We moved the position of the vanadium foil, starting \qty{0.5}{\milli\meter} downstream from the nominal focus, and moved four \qty{250}{\micro\meter} steps upstream. The foil thickness was much smaller than the nominal Rayleigh length of approximately \qty{200}{\micro\meter} as indicated in \autoref{fig:experiment}~(b) and (c). Thus, the scan produced a direct map of the multi-shot-averaged focal volume, which is difficult to access by other means due to the FEL's spatial jitter without significant attenuation \cite{classen2017, nakamura_focus_2020, trost2023imaging, sikorski2015, Sun2021contrast}. The FEL generated a fluorescing needle with the base formed by the Gaussian-line FEL focal spot. The thickness of the foil determines the length of the needle.
%%%%%%%%%  Figure 3: Experimental results  %%%%%%%%%%%
We mapped both the 3D structure of the fluorescing needle and the focal spot changes due to an astigmatism by scanning the sample through the focus. We used intensity correlations from several thousand exposures per foil position to determine the $g^{(2)}$ functions for each scanned FEL focal-volume slice.
At the nominal focus position of the foil \gtwo{}  functions calculated from 16 detector tiles (\autoref{fig:results}~(a)) mirror different projections of the fluorescing needle (\autoref{fig:results}~(b)). 
 The central tiles observe a 2D projection of the focal spot point on. The outer tiles show elongated and rotated \gtwo{} patterns depending on the viewing angles relative to the long axis of the needle up to $\approx \qty{9}{\degree}$.  
The focal scan of the foil position provides a clear map of the beam's astigmatism (\autoref{fig:results}~(c)). The shape of the central projection evolves from a round spot at the focus to an elongated ellipse further up or downstream. The axis of ellipticity flips by \qty{90}{\degree} across the focal plane. This is a distinct signature of an astigmatism (cf. \autoref{fig:results}~(d)).
We fitted the FWHM of the main axes x and y of the ellipses using the $g^{(2)}$ and assuming a Gaussian intensity distribution inside the FEL focus spot. A description of the fitting procedure is given in the Supplemental Material.  Measurements fluctuate between different runs, but the overall focus spot size is larger than the nominal focal spot size of $< \qty{200}{\nano\meter}$. From the tilt of the major axis of the ellipses (as visualized in \autoref{fig:results}~(d) by the orange arrow), one can also estimate the mismatch between the angle of incidence on the vertical and horizontal KB mirrors. 
The highest measured value for $g^{(2)}$ is 1.03, which corresponds to pulse durations $<\qty{3}{\femto\second}$ on average \cite{inoue_determination_2019,trost_speckle_2023}. In principle, the non-linear compression scheme can deliver sub-fs pulses, however, in the present case, most pulses contained multiple spikes, which increase the measured FEL pulse duration. 
Using IDI, we were able to image the astigmatic focal profile and simultaneously the FEL pulse duration. The measurement required only 5,000–10,000 exposures per sample position, despite significant FEL spatial jitter. The entire data acquisition takes less than 20~minutes at $\qty{120}{\hertz}$, making IDI a practical tool for in-situ monitoring without the need to attenuate the beam.
%%%%%%%%%%%%%%%%  Figure 4: Experimental vs Simulation 3D reconstruction %%%%%%%%%%%
Each distinct sample position along the astigmatism will result in a different shape of the fluorescing needle, as illustrated in \autoref{fig:simulation}~(a). We compare the experimentally measured \gtwo{} functions with the simulation of fluorescing \qty{4}{\micro\meter} needles using the bases extracted from \autoref{fig:experiment}~(c) and (d). The simulation assumes that photons are emitted with random phases. The amplitudes of the electric field are propagated into the far-field, where the interference pattern is measured as a speckle pattern. The \gtwo{} functions for each tile are calculated from the intensity correlations. Overall, the simulated and experimental \gtwo{} functions are in good agreement. The experimental data indicate that there are some deviations from the Gaussian focus assumptions. Some of the tiles show faint focal wings, which would not appear in the case of a perfect Gaussian beam.  

%%%%%%%%%%%%%%%%%%%%%%%%   Discussion     %%%%%%%%%%%%%%%%%%%%%%%%%%%%%%%
\section{Conclusion and outlook}
We have demonstrated that IDI can be used to obtain multiple projections of X-ray fluorescing nanoscale objects without the need to rotate the specimen.. In contrast to CDI, the recorded projection of the sample depends on the detector position, rather than the direction of the illuminating X-ray pulse. The contrast of the projection can be further increased by shortening the FEL pulses. We estimate that the FEL pulse duration in this experiment was 2-3 fs short based on calculations presented in \cite{inoue_determination_2019,trost_speckle_2023}. Such few-fs pulse durations are up to three times longer than the life time of the vanadium K$\alpha$ line. Newly available sub-fs hard X-ray pulses can in theory increase the contrast of \gtwo{} by an order of magnitude.

Potential future directions for IDI include integrating IDI with coherent diffractive imaging to acquire multiple specimen projections in a single orientation \cite{ho2021fluorescence}, combining IDI with spectroscopic measurements such as resonant inelastic X-ray scattering or Raman spectroscopy  \cite{Wernet2019chemical, wollweber_nanoscale_2024, Tamasaku2023twoD-Kb-kalpha, Inoue2021shortening}, and precise studies of non-linear sequential multiphoton absorption \cite{Tamasaku2023twoD-Kb-kalpha, Kuschel2025non, ho2021fluorescence}.
For the combination of IDI and CDI, even low-resolution projections of nanoscale specimens can serve as three-dimensional constraints for the reconstruction of CDI images. IDI can also increase the material specificity of CDI and might reduce the number of exposures required to obtain precise 3D maps of nanoscale specimens. Under ideal circumstances, IDI could, in theory, achieve atomic resolution for specimens consisting of only few spatial frequencies. However, the required signal-to-noise ratios remain subject to debate \cite{schneider2018quantum, trost_speckle_2023, lohse_incoherent_2021}. One limitation is nanoplasma formation during the imaging process, which broadens the emitted fluorescence spectrum. However, this can be partially overcome by using fluorescence from transient resonances, which have narrow bandwidths and only arise at very high intensities \cite{Tamasaku2023twoD-Kb-kalpha, Kuschel2025non}.
Combining three-dimensional IDI with spectroscopic measurements would enable the integration of ionic structural information with spectroscopic fingerprints of electronic processes, such as charge transfer, in a single measurement. The broad-band nature of short X-ray pulses required for IDI can be overcome through stochastic data analysis, such as recently demonstrated, in super-resolution stimulated X-ray Raman spectroscopy \cite{Li2025super}.
IDI of fluorescence triggered by non-sequential multiphoton absorption could be used to avoid focal intensities averaging effects in studies of non-linear X-ray-matter interaction. Often, the non-linear signal originates from a small focus fraction, with the highest intensities near the focus center, and is overlapped with the signal from the significantly larger focal wings. This can significantly obscure non-linear signatures as has been demonstrated in the past \cite{bostedt2012ultrafast, Gorkhover2012, ferguson2016transient}. IDI can potentially map the origin of non-linear processes in 3D and remove focal averaging effects.
In principle, the measurement of \gtwo{} can be added to most imaging or spectroscopic experiments parasitically. We conducted a cross experiment at SACLA \cite{ishikawa2012}, which demonstrates that \gtwo{} can be recorded even in a perpendicular geometry, which can be easily upgraded by an additional forward scattering detector or even an in-line spectrometer. See the Supplemental Material for a description of the experiment at SACLA.

In summary, IDI is an interesting new route to combine three-dimensional structural information with spectroscopic fingerprints without the need to rotate the sample. Our findings may be relevant for the study of non-equilibrium states of matter, such as chemical reactions, phase transitions, and fundamental light-matter interactions. 

\clearpage

\section{Contributions}

T.G. conceived the idea for the experiment based on discussions with S.K. and I.I.; A.M. implemented TW sub-fs to few-fs pulses. All authors contributed to the experimental campaign at LCLS and SACLA. R.R. carried out data analysis with input from T.G., S.K. and other authors. F.Z. wrote and carried out the simulation. R.R, F.Z. and T.G. wrote the manuscript with the input from all authors.

\begin{acknowledgments}
We would like to acknowledge the supporting members of the LCLS and SACLA facility.
Use of the Linac Coherent Light Source (LCLS), SLAC National Accelerator Laboratory, is supported by the U.S. Department of Energy (DOE), Office of Science, Office of Basic Energy Sciences (BES) under Contract No. DE-AC02-76SF00515.
The second experiment was performed at SACLA with the approval of the Japan Synchrotron Radiation Research Institute (JASRI) (proposal nr. 2019B8072).
T.G. acknowledges funding by the European Union (ERC Starting Grant 101040547 - HIGH-Q). Views and opinions expressed are however those of the author(s) only and do not necessarily reflect those of the European Union or the European Research Council Executive Agency. Neither the European Union nor the granting authority can be held responsible for them. A.U. acknowledges funding by the Cluster of Excellence
``CUI: Advanced Imaging of Matter'' of the Deutsche Forschungsgemeinschaft (DFG)
– EXC 2056 – project ID 390715994.
A. M. acknowledges funding from the U.S. Department of Energy, Office of Basic Energy Sciences, Accelerator and Detector research program. C.H. and R.Z. acknowledge funding by the Deutsche Forschungsgemeinschaft (DFG) within the Collaborative  Research Initiative SFB 986 ``Tailor-Made Multi-Scale Materials Systems'' (project number 192346071).
The authors thank Rasmus Buchin for fruitful discussion.
\end{acknowledgments}

\bibliography{library}% Produces the bibliography via BibTeX.
\clearpage

\printfigures
\appendix
\section{Supplemental Material}

\subsection{Data processing}

\begin{figure}[p]
    \centering
    \includegraphics[width=\textwidth]{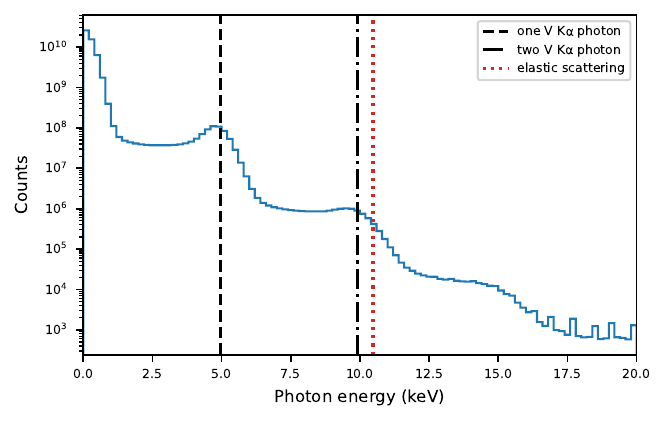}
    \caption{
    Histogram of the detected photon energies over 24000 shots. The black lines mark the energies of one (dashed) and two (dash-dot) detected vanadium $\text{K}_\alpha$ photons at \qty{4.95}{\kilo\eV}. The red line marks the energy of potential elastically scattered photons at \qty{10.48}{\kilo\eV}. No significant number of elastically scattered photons is observed.
    }
    \label{fig:histogram}
\end{figure}

Data processing was performed using resources of the SLAC Shared Science Data Facility (S3DF).
First, the energy-resolved, calibrated data from the Jungfrau~4M detector were converted to photons pixel-wise by thresholding. Faulty pixels and pixels receiving scattered light at the very center of the detector were masked. A histogram of the calibrated Jungfrau~4M data is shown in \autoref{fig:histogram}. The absence of a feature at the energy of the incoming radiation at \qty{10.48}{\kilo\eV} (red dotted line) indicates that notable contamination of the fluorescence signal by elastically scattered photons is unlikely. For the analysis, every $1024 \times 512 \;\text{pixel}$ module of the Jungfrau~4M was split into eight individual $256 \times 256\;\text{pixel}$ tiles. These tiles have an angular extent of about \qty{2}{\degree}.
For each tile, the data were then filtered by the number of fluorescence photons recorded. Only shots with at least 1000 photons on the respective tile were used to calculate the second-order correlation. 1000~photons amounts to \qty{1.5}{\percent} of the tiles pixels recording a photon. This way, 4000 to 11000 single images were processed for every tile and averaged to obtain the average second-order correlation per tile. The second-order intensity correlation of all four upper and lower tiles of every detector module was then averaged to yield the bigger tiles shown in \autoref{fig:results} and \autoref{fig:simulation}. The code used for the second-order calculation is available at \cite{kuschel2022}.

\subsection{Fitting procedure}

We approximate the fluorescing volume created on the vanadium foil by a two-dimensional Gaussian. Employing a phase-retrieval algorithm is unnecessary for such a simple object; instead, we can directly fit the expected Fourier amplitudes. We start from the approximate spatial distribution of the fluorescence $S(x,y)$ given by
$$S\left(x,y\right) = \frac{1}{
2 \pi  \sigma_{x} \sigma_{y} \sqrt{-\rho^{2}+1}
}\exp\left[
    \frac{
        \sigma_{x}^{2} y^{2}+\sigma_{y}^{2} x^{2} -2 \rho \sigma_{x} \sigma_{y} x y
    }
    {
        2 \sigma_{x}^{2} \sigma_{y}^{2} \left(\rho-1\right) \left(\rho+1\right)
    }
\right]
$$
Where $\rho$ is the correlation coefficient of the Gaussian and $\sigma_{x}$ and $\sigma_{y}$ are the standard deviations in the $x$ and $y$ direction, respectively, assuming no correlation between the $x$ and $y$.
To obtain the structure factor of the fluorescence-emitting area, we take the two-dimensional Fourier transform of $S(x,y)$, yielding
% Normalized
$$ \left(\mathcal{F}S\right)(q_x,q_y) = \tilde{S}(q_x,q_y) = 
\exp\left[
    -\frac{1}{2} q_{x}^{2} \sigma_{x}^{2}
    -\frac{1}{2} q_{y}^{2} \sigma_{y}^{2}
    -q_{x} q_{y} \rho \sigma_{x} \sigma_{y}
\right]
$$
As $g^{(2)}(q_x,q_y) = 1 + \beta\left|\tilde{S}{q_x,q_y}\right|^2$ is related to the absolute square of $\tilde{S}(q_x,_qy)$ the final fit function is given by
% Magnitude squared
$$\beta\left|\tilde{S}(q_x,q_y)\right|^2 = 
\beta\exp\left[
    - q_{x}^2 \sigma_{x}^2
    - q_{y}^2 \sigma_{y}^2
    -2 q_{x} q_{y} \rho \sigma_{x} \sigma_{y}
\right]
$$
with $\beta$ being the visibility factor of the second-order described in greater detail in \cite{inoue_determination_2019, trost2023imaging}.
The non-linear regression of all parameters was performed using a standard Levenberg-Marquardt algorithm as provided by \cite{matt_newville_2024_12785036}.
\subsection{Simulation\label{chap:simulation}}
The IDI speckle patterns used to calculate \gtwo{} in \autoref{fig:simulation} were simulated by modeling discrete emitters located at random positions within the vanadium foil\cite{masterthesis,zenodo}. Their spatial probability distribution matched the astigmatic Gaussian beam profile obtained from our experimental fits. For each shot, the phase of each emitter was randomized (\(\phi_n\sim U(-\pi,\pi)\)), and the intensity at a detector location \(\vec{x_i}\) was calculated as
$$
I(\vec{x_i})\propto\left|\sum_n^N \frac{1}{\left|\vec{r}_n-\vec{x_i}\right|} e^{i(k{\left|\vec{r}_n-\vec{x_i}\right|}+\phi_n)}\right|^2
$$
where \(k\) is the Vanadium K\(_\alpha\) wavenumber (\(k=\qty{25}{\per\nano\metre}\)). Each \gtwo{} function was then determined by averaging intensity correlations over 100 shots.
\subsection{\label{chap:SACLA} Cross experiment at SACLA}

\begin{figure}[p]
    \centering
    \includegraphics[width=\textwidth]{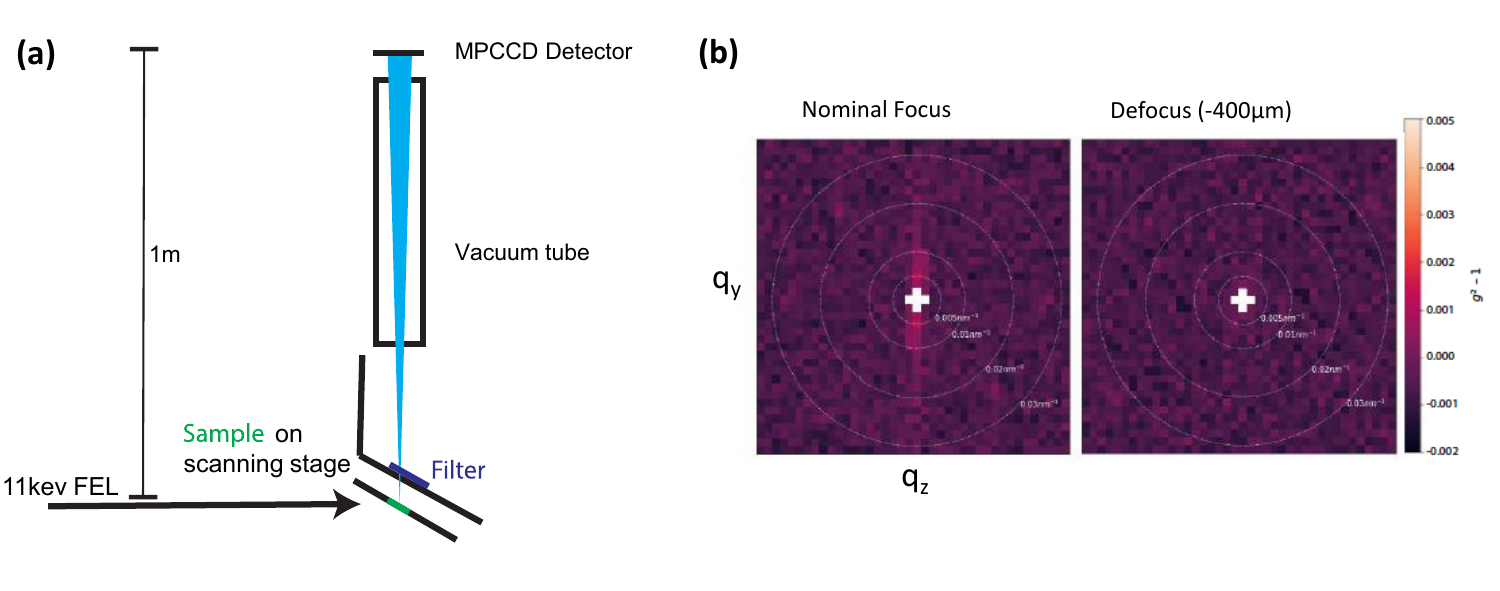} 
    \caption{
    Additional experimental data measured at SACLA FEL.
    \textbf{a}~The experimental setup at SACLA consisted of a \qty{10}{\micro\meter}-thick Fe foil and a dual MPCCD detector placed perpendicular to the \qty{11}{\kilo\eV} X-ray beam.
    \textbf{b}~Measured \gtwo{} on the Fe foil at the nominal focus and \qty{400}{\micro\meter} upstream.
    }
      \label{fig:SACLAIDI}
\end{figure}

We have conducted a complementary experiment at SACLA \cite{ishikawa2012}. The experimental setup is illustrated in panel (a) of \autoref{fig:SACLAIDI}. Here, the mpCCD detector \cite{kameshima2014} observed the fluorescing volume perpendicular to the incoming FEL. The volume consisted of a \qty{10}{\micro\meter} thin Fe foil intersected by nanofocused FEL pulses with durations around 6-\qty{8}{\femto\second}. The Fe foil was rotated by \qty{45}{\degree} relative to the mpCCD detector and the FEL direction. The distance between the mpCCD detector and the fluorescing volume was 1 m. We attenuated the X-ray beam by a \qty{0.4}{\milli\meter} thin Si wafer in order to avoid non-linear effects.  

Wire scans confirmed a FWHM of \qty{200(10)}{\nano\meter} in y-direction, which is perpendicular to the FEL and the x-axis. We calculated \gtwo{} for 5000 shots in nominal focus (left, panel (b)) and outside of the focus (right, panel (b))\cite{zenodo}. The fit routine provides a nominal focus FWHM estimation of \qty{240(20)}{\nano\meter}, which is close to the wire scan value, considering the elongation due to the 45-degree tilt of the foil. The z-direction is significantly undersampled due to the thickness of the foil. In the defocused position, the focus is undersampled in both directions as expected.

Overall, this measurement demonstrates the 3D imaging capabilities of IDI in a setup that is compatible with forward scattering and spectroscopy experiments.

\end{document}